\documentclass[prb,twocolumn,showpacs]{revtex4}
 \usepackage{graphicx}
 \usepackage{dcolumn}

\begin{document}

 \title{Effect of hydrogen on the atomic structure of Pd(001)}

 \author{S.H. Kim, J. Barthel, H.L. Meyerheim, and J. Kirschner}
  \affiliation{Max-Planck Institut f\"ur Mikrostruktur Physik,
 Weinbergweg 2, Halle/Salle, Germany }
 
 \author{Jikeun Seo}
 \affiliation{Department of Ophthalmic Optics, Chodang University,
  Muan 534-701, Korea }
 
 \author{J.-S. Kim}
 \affiliation{Department of physics, Sook-Myung Women's university,
  Seoul 140-742, Korea}
 \date{\today}

 \begin{abstract}

  The atomic structures of clean and hydrogen-adsorbed Pd(001) are
 investigated by low energy electron diffraction (LEED) I/V analysis. Clean
 Pd(001) shows little surface relaxation in sharp contrast to previous reports.
 Adsorbing 1 monolayer of hydrogen on Pd(001), we observe sizable
 expansion of the interlayer spacing of the first two surface layers, $d_{12}$
 by 4.7\% of the corresponding one of bulk Pd. Both experimental
 observations are in excellent agreement with the predictions of recent {\it
 ab initio} calculations. A series of experiments with varying coverages of
 hydrogen adsorbed on Pd(001), reveals that $d_{12}$ monotonically increases
 with the increasing coverage. Such an observation strongly supports the
 contention that the previous observation of expanded $d_{12}$ in clean
 Pd(001) results from contamination of the surface by residual hydrogen.
 \end{abstract}
 \maketitle

 \section{introduction}

  Hydrogen is well known to diffuse and dissolve well in Pd, which has
 motivated researchers from many different disciplines to investigate Pd-H
 complex due to its possible application as hydrogen carrier.\cite{Hydrides}
 The adsorption, dissociation and diffusion of hydrogen on single crystalline
 Pd has been studied as a model system to understand atomistic process of
 hydrogen incorporation and catalytic behavior of Pd. For Pd(001), numerous
 experimental works have been performed by employing low energy electron
 diffraction (LEED) \cite{Behm0,Jona}, He ion scattering\cite{Rieder}, work
 function measurement\cite{Behm}, thermal desorption spectroscopy
 (TDS)\cite{Behm0,Okuyama,Wilde}, nuclear reaction analysis
 (NRA)\cite{Besenbacher,Wilde}, electron energy loss spectroscopy
 (EELS)\cite{Nyberg,Okuyama}. Moreover, {\it ab initio} calculations have
 been made to study electronic and atomic structures, and chemical
 properties of Pd(001) upon hydrogen
 adsorption.\cite{Tomanek,Wilke1,Wilke2,Eichler} Thereupon, found are many
 physical and chemical properties of Pd(001) such as dissociation mechanism
 of $H_2$, adsorption sites and adsorption energies of hydrogen, dependence
 of work function on the coverage of hydrogen, desorption and dissolution
 kinetics of hydrogen on Pd(001).\par

  The atomic structure of clean and hydrogen-adsorbed Pd(001) is,
 however, not established yet. Behm $et. al.$\cite{Behm0} and Jona $et.
 al.$\cite{Jona} found from their LEED I/V analysis that the interlayer
 spacing of the first two layers from the surface ($d_{12}$) expanded by 
 2.5 \%, 3.0 \%, respectively, with respect to the coresponding bulk spacing,
 $d_B$. Most unreconstructed metal (001) surface, however, show
 contraction of $d_{12}$, due to increased bond strength between the two
 layers resulting from the redistribution of electrons from surface atoms to
 the first interlayer space.\cite{Feibelman} In this regards, the expansion
 of $d_{12}$ of Pd(001) is an unexpected observation.\par

   As the origin for such an expansion of $d_{12}$, Jona $et. al.$\cite{Jona}
 pointed out two possibilities; magnetization of the surface and hydrogen
 contamination of Pd(001). Magnetization of Pd thin ($<$ 5ML) films grown on
 Ag(001)\cite{Erskine} and Au(001)\cite{Bader} was investigated by MOKE,
 but null magnetization is found.  Both self-consistent tight binding
 calculation\cite{Demangeat} and {\it ab initio} calculation by full potential
 liniearized augmented plane wave method (FLAPW) predicted delicate
 dependence of magnetization of Pd on the thickness of slab to model Pd(001).
({\it Note: at the moment we are waiting for the result of Prof. Hong 
on the effect of magnetization on surface relaxation pf Pd(001).
He suggested  very small, if any, relaxation due to the tiny magnetization of Pd.})
 On the other hand, for hydrogen covered Pd(001),
 expansion of $d_{12}$ is predicted.\cite{Tomanek,Wilke1,Eichler} 
 However, no direct experimental
 investigation has been made yet to examine the correlation between
 hydrogen coverage and the atomic structure of Pd(001). Such study should
 be critical to identify the origin of the observed expansion of $d_{12}$.\par

  The present work is aimed to experimentally clarify the effect of hydrogen
 coverage on the atomic structure of Pd(001). We carefully prepare Pd(001)
 with various hydrogen coverages, and investigate their atomic structure by
 LEED I/V analysis. From that study, we find that there is little
 expansion of $d_{12}$ for clean Pd(001), while the $d_{12}$ monotonically
 expands with increasing hydrogen coverage. We also find that the remnant
 hydrogen in our experimental chamber, although its base pressure is very
 low, low $10^{-11}$ mbar, can swiftly contaminate Pd(001) and result in
 the notable expansion of $d_{12}$. Combining all the abovementioned
 observation, we conclude that expansion of $d_{12}$ of clean Pd(001)
 previously reported by other experimental groups\cite{Behm0,Jona},
 originates from hydrogen contamination of Pd(001).\par

  \section{Experiments and LEED I/V analysis}

  \subsection{Experiment}

  All the experiments were performed in an ultrahigh vacuum chamber with
 its base pressure low $10^{-11}$ mbar. The chamber was equipped with
 rear-view LEED optics and cylindrical mirror analyzer for Auger electron
 spectroscopy (AES). Pd(001) sample was of "top-hat" shape with its
 diameter, 8 mm, and thickness, 1 mm. We cleaned the sample by iterating
 sputtering with 2 KeV Ar$^+$ and annealing at 950 K for 20 minutes. To
 remove remnant carbon, we oxidized it by annealing the sample at 650K in
 an oxygen ambient pressure of $1\times10^{-8}$ mbar, and desorbed it at
 950 K, until there was observed no AES peak of carbon. Once the
 contaminants were removed of the sample, we prepared the sample by
 sputtering and annealing only for once. From now on, we call this routine
 sample preparation procedure, $standard$ procedure.\par

  For the $standard$ procedure, the time required to complete acquisition of
 I/V spectra was around 2 hours from the start of annealing. During that
 period, it seemed highly possible that the surface was contaminated by
 residual hydrogen. Thus, we prepared clean sample in two different ways
 to ensure cleanliness of Pd(001). The first way was to minimize the time
 till I/V acquisition; After following the $standard$ procedure, we flashed
 the sample up to 950 K to detach any residual hydrogen and cooled the
 sample to 150 K within 30 minutes by liquid nitrogen. ( Desorption
 temperatures of surface and subsurface hydrogen, $T_D$, were known to
 be lower than 340 K. \cite{Okuyama}) Following this $rapid$ cleaning
 process, we could finish the I/V acquisition within 1 hour in the better
 vacuum condition. The second method was based on the fact that the
 desorption temperature of hydrogen was around 340 K; We took I/V with
 the sample at temperatures higher than the desorption temperature. \par

  Hydrogen covered Pd(001) was also investigated as follows; After flashing
 the sample at 950 K, we cooled it down to 150 K for hydrogen dose,
 because the hydrogen was known to dissolve into bulk Pd near room
 temperature. \cite{Wilde,Okuyama} We dosed 6$\sim$12 Langm\"ur (L) of
 hydrogen in an ambient hydrogen pressure, $1\times10^{-8}$ mbar. Since
 there was little difference in LEED I/V for dosing more than 6 L of
 hydrogen, we concluded that 6 L of hydrogen was enough to saturate
 Pd(001). Hydrogen saturated Pd(001) showed $p(1\times 1)$ LEED pattern
 with its background intensity about the same as for clean Pd(001) on
 visual inspection.\par

  \subsection{LEED I/V Analysis}

  LEED I/V spectra were obtained by a fully automated video-LEED system
 comprised of a charge coupled device (CCD) camera and a program for
 image processing. The I/V spectra were always taken with the sample
 normal to the electron beam. LEED I/V analysis was made by SATLEED
 program.\cite{tleed} Scattering phase shifts were obtained from potentials
 of Moruzi $ et. al.$\cite{moru} for its angular momentum, $l$ from 0 to
 10. Thermal vibration effect was taken into account by Debye-Waller factor
 with Debye temperatures of H and Pd, 1800 K and 260 K, respectively.
 We also varied the Debye temperatures as fitting parameters, however, the
 best-fit atomic structure was not sensitive to them. The quality of I/V fit
 was judged by the reliability factor of Pendry, $R_P$, and error limit was
 set by its variance.\cite{variance} Fitting was made for various model
 structures, and the best-fit structure was concluded after iterating the
 fitting till the difference between the input structural parameters and
 those of the resulting best-fit structure were within 0.001 \AA. \par

  \begin{table*}
  \caption{Results of LEED I/V analysis for $rapidly$ cleaned Pd(001).
 {\it Clean}, {\it Hollow-H}, {\it Subsurface-H} respectively signifies model structures,
 clean Pd(001),  Pd(001) with 1ML hydrogen in the four-fold
 hollow site of the surface layer, and  Pd(001) with 1 ML of hydrogen 
 in the four-fold hollow site beneath Pd surface layer.}
  \begin{ruledtabular}
  \begin{tabular}{ccccccc}
  & \multicolumn{3}{c}{Model structure} &
  \multicolumn{2}{c}{Theory} & Experiment \\ \cline{2-4} \cline{5-6}
  & {\it Clean} & {\it Hollow-H} & {\it Subsurface-H} & LDA\cite{Wilke1} &
 GGA\cite{Eichler}
    & (Jona)\cite{Jona} \\
  \tableline
  $\Delta d_{12}/d_B$ & {\bf +0.2}$\pm$1.4 \% & -0.4 \% & -0.4 \%
       & -0.6 \% & -1.0 \% & +3.0 \% \\
  $\Delta d_{23}/d_B$ & {\bf -0.7}$\pm$1.3 \% & +0.0 \% & +0.3 \% &
 & +0.1 \% & -1.0 \% \\
  $R_P$-factor & {\bf 0.1705} & 0.1944 & 0.1941 &  & & 0.350 \\
  variance &  0.0206 &  &  & & & \\
  \end{tabular}
  \end{ruledtabular}
  \end{table*}

  \section{Results and Discussion}

  \begin{figure}
  \includegraphics[width=0.45\textwidth]{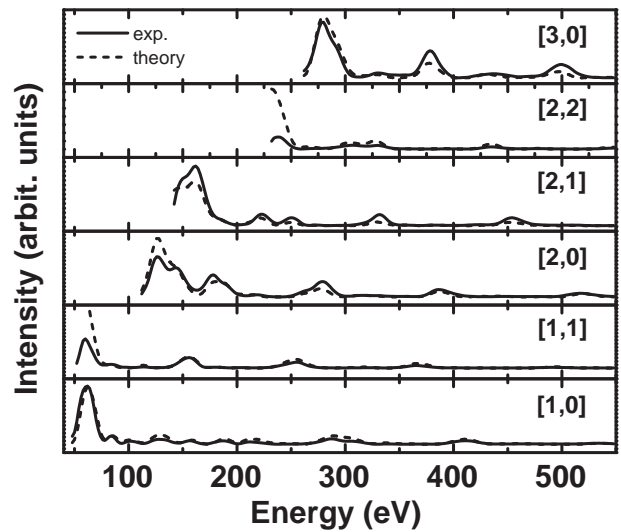}
  \caption{ LEED I/V spectra for $rapidly$ cleaned Pd(001).
 Thick line is experimental curve and the dotted line is theoretical one.}
  \end{figure}

  In Fig. 1, shown are LEED I/V spectra for [1,0], [1,1], [2,0], [2,1], [2,2],
 [3,0] beams of the $rapidly$ cleaned Pd(001). The total energy range of the
 spectra is 2468 eV. In Table I, summarized are the best-fit structures for
 three, most conceivable models, given the present experimental condition
 and the results in literatures. The minimum R-factor is found for clean
 Pd(001) model and is very small, 0.17, in regards to the large number of
 beams and extensive energy range of I/V spectra. The theoretical I/V
 spectra of the best-fit structure in Fig.1 well reproduces most features in
 the experimental spectra, which purports the present I/V fitting reliable.
 $R_P$-factors for the other models, some of which are given in Table I, are
 outside the variance of the minimum $R_P$-factor of clean Pd(001), and
 thus they are excluded for the model of $rapidly$ cleaned Pd(001). \par

  The best-fit structure for $rapidly$ cleaned Pd(001) shows little surface
  relaxation (Table I); $d_{12}$ shows a negligible amount of expansion by
 + 0.2 \% of $d_B$. (From now on, all structural variation is given with
 respect to $d_B$.) $d_{12}$ for clean Pd(001) is predicted to contract by
 less than 1\% by recent theoretical works ignoring 
 spin-polarization\cite{Eichler,Methfessel}, which is in
 good agreement with the present experimental result. Hence, it is likely
 that the $rapidly$ cleaned sample is really clean Pd(001). \par

  \begin{table*}
  \caption{ Results of LEED I/V analyses for H-adsorbed Pd(001) at 150 K.
 $d_{H}$ and $d_{12}$ are as defined in Fig. 2. Among others, four most
 relevant model structures, Pd(001) with hydrogen at fourfold hollow site,
 {\it Hollow-H}, {\it Bridge-H}, {\it On-top -H}, and {\it Clean} signify structures where 1 ML
 of hydrogen is adsorbed respectively on four-fold hollow site, bridge site,
 and on-top site, and clean Pd(001).}
  \begin{ruledtabular}
  \begin{tabular}{cccccccc}
  & \multicolumn{4}{c}{Model structure} &
 \multicolumn{2}{c}{Theory}  & Experiment\cite{Besenbacher} \\ \cline{2-5} \cline{6-7}
     & {\it Hollow-H} & {\it Bridge-H} & {\it On-top-H} & {\it Clean}
   & LDA\cite{Wilke1} & GGA\cite{Eichler} &  \\
  \tableline
  $d_{H}$  & {\bf 0.20 }$_{-0.22}^{+0.43}$ \AA & 1.21 \AA & 1.10 \AA &
     & 0.16 \AA(hollow) & 0.20 \AA & 0.3 \AA \\
  $\Delta d_{12}/d_b$ & {\bf +4.7}$\pm$1.1 \% & +5.2 \% & +5.8 \% & +5.3 \%
     & +5.2 \% & +4.4 \% & ($d_{H}+\Delta d_{12}$) \\
 $\Delta d_{23}/d_b$ & {\bf +0.0}$\pm$0.9 \% & -1.0 \% & -0.1 \% & -0.4 \%
       &   & +0.2 \% & \\
  R-factor & {\bf 0.1626} & 0.2720 & 0.3594 & 0.2646 & & & \\
  variance &  0.0231 &  &  &  & & & \\
  \end{tabular}
  \end{ruledtabular}
  \end{table*}

  \begin{figure}
  \includegraphics[width=0.45\textwidth]{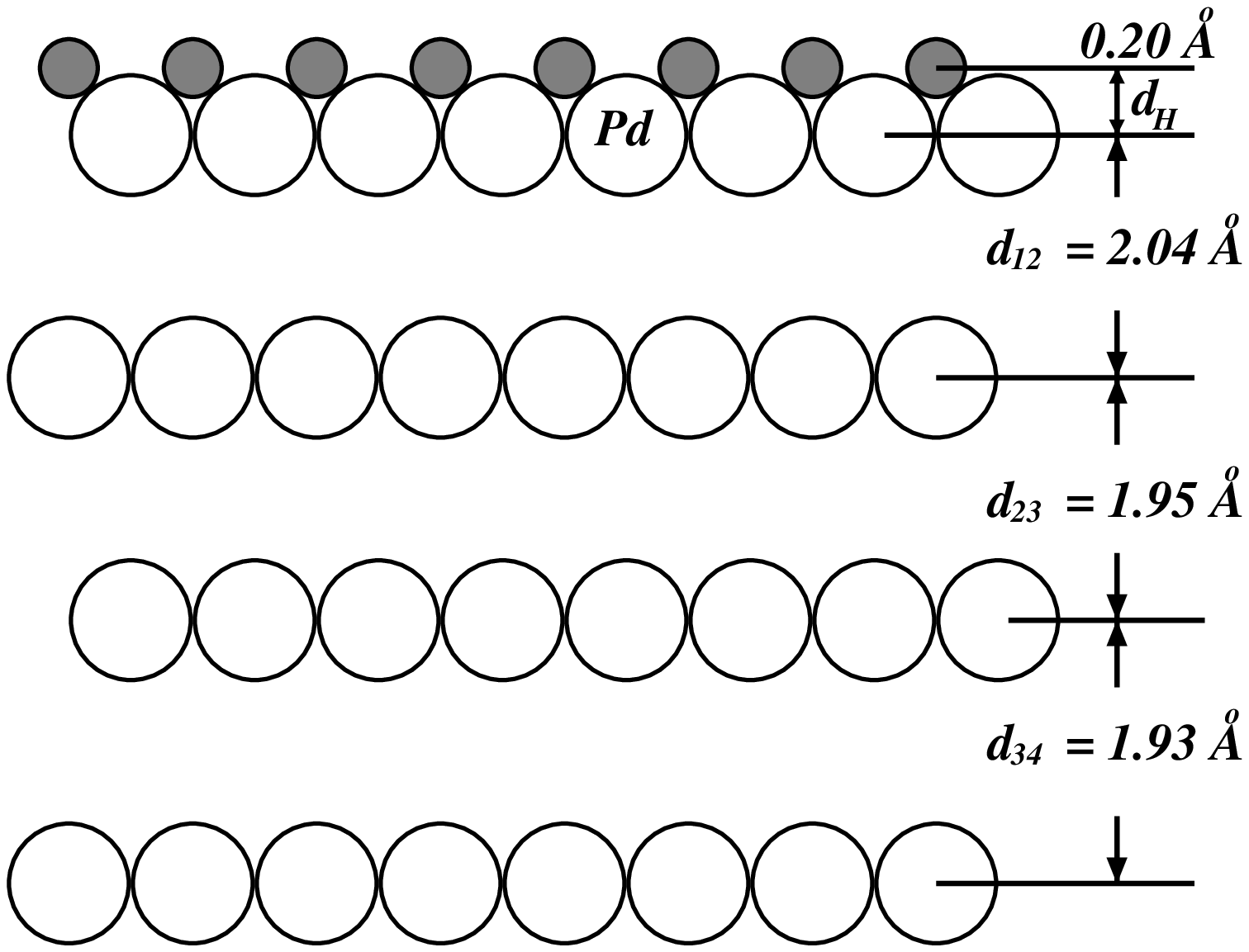}
  \caption{The best-fit structure of nominal 1 ML of hydrogen covered
  Pd(001).}
  \end{figure}

  \begin{figure}
  \includegraphics[width=0.45\textwidth]{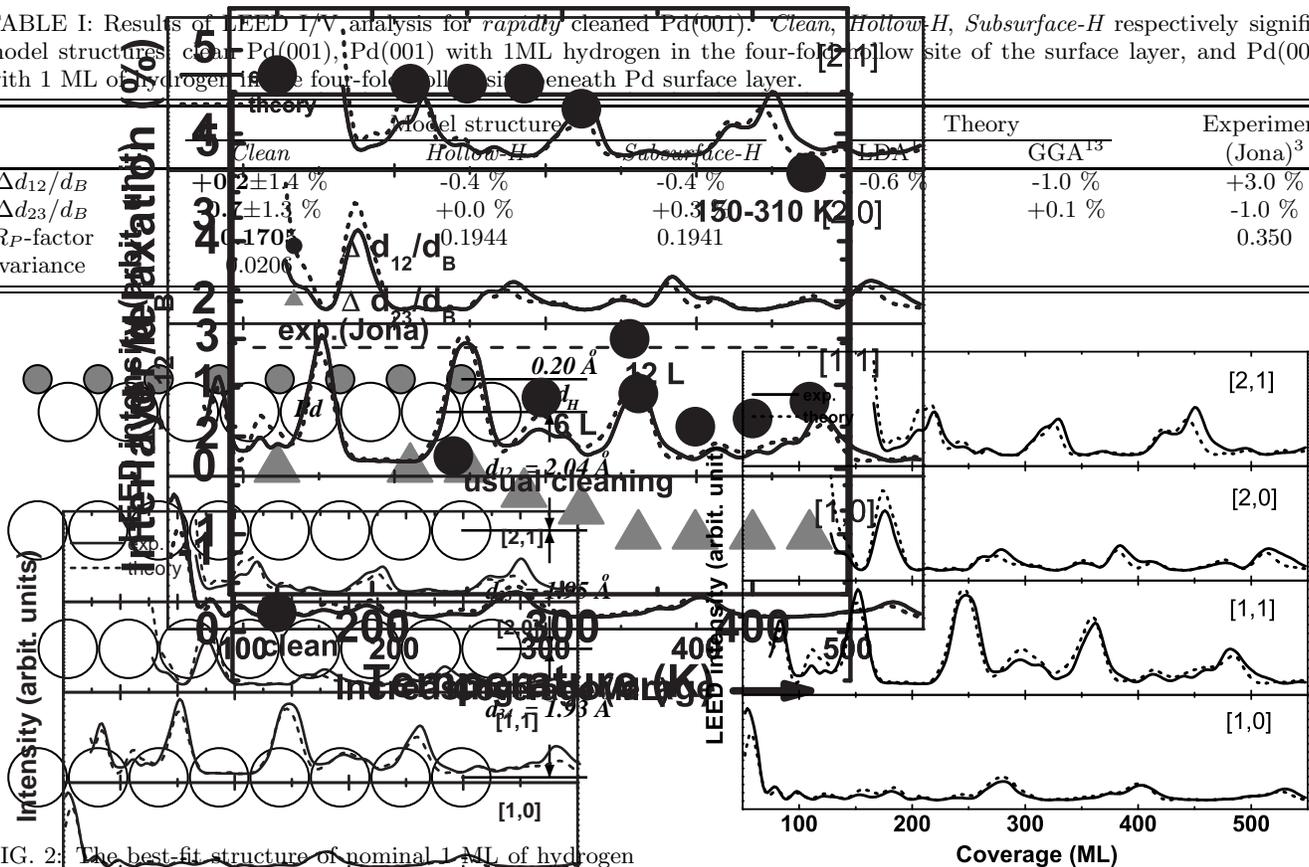}
  \caption{ LEED I/V spectra for the nominal 1ML of hydrogen covered
 Pd(001). Solid line signifies theoretical result and the doted one does
 theoretical spectra for the best-fit structure.}
  \end{figure}

  The atomic structures of both hydrogen-adsorbed and subsequenly
 hydrogen-desorbed Pd(001) are also investigated by LEED I/V analyses.
 The I/V spectra are obtained after dosing 6 L of hydrogen on Pd(001) at
 150 K and then while gradually warming up the sample beyond $T_D$.
 Detailed sample preparation procedure is as described in the experiment
 section.\par

  In the beginning, the atomic structure of hydrogen-adsorbed Pd(001) at 150
 K is investigated by the analyses of LEED I/V spectra for [1,0],
 [1,1],[2,0],[2,1] beams with their total energy range, 1810 eV. Table II is a
 summary of the I/V analysis. The best-fit is found for a model where 1
 monolayer (ML) of hydrogen is adsorbed on the four-fold hollow site of
 Pd(001).(Fig. 2) We find that the theoretical I/V spectra for the best-fit
 structure well reproduce all the experimental features in Fig. 3. The
 $R_P$-factor of the best-fit structure is also quite low, 0.1626, and the
 other structures result in $R_P$-factors far beyond the variance of the
 minimum $R_P$-factor (Table II), assuring that the best-fit structure
 represents the atomic structure of the hydrogen-adsorbed Pd(001).\par

  For the best-fit structure (Fig.2 and Table II), the distance between
 hydrogen and Pd surface layer, $d_H$, is 0.2 \AA. Large error limit of
 $d_H$ reflects small scattering cross section of hydrogen. $d_{12}$ expands
 by 4.7 \%, and $d_{23}$ and $d_{34}$ show no relaxation. The present
 observation is in excellent agreement with recent theoretical predictions that
 four-fold hollow site is the energetically most favored adsorption site of
 hydrogen till its coverge reaches 1 ML, and $d_H$ is 0.1 $\sim$ 0.25 \AA, 
 and $d_{12}$ expands by 4.4 $\sim$ 5.2 \%.
 \cite{Tomanek,Wilke1,Eichler} Besides, ion-channeling
 experiment\cite{Besenbacher} predicts $d_H$ to be 0.3 \AA. In this
 experiment, $d_H$ includes $\Delta d_{12}$, the variation of $d_{12}$
 from $d_B$. If we also include $\Delta d_{12}$ to $d_H$, the resulting
 value is 0.29 \AA , in excellent agreement with the result of
 ion-channeling experiment, 0.3 \AA. \par

  We trace the evolution of its surface structure by LEED I/V analysis
 while gradually annealing the sample to temperatures higher than $T_D$,
 since we expect to obtain clean Pd(001) by desorbing hydrogen from the
 sample. In Table III, below 340 K, the best-fit model is Pd(001) with 1
 ML of hydrogen adsorbed on four-fold hollow site. Further, $d_{12}$ and
 $d_{23}$ are maintained around the same value as that at 150 K. If the
 sample temperature ($T_S$) is raised higher than $T_D$, the best-fit model
 switches from hydrogen-adsorbed Pd(001) to clean Pd(001) (Table III), and
 both $d_{12}$ and $d_{23}$ simultaneously show such abrupt contraction
 as demonstrated in Fig. 4. Although the $R_P$-factors of both structures
 are within the variance of the minimum R-factor, such an observation of
 the change of the best-fit structure is consistent with the observation of
 hydrogen desorption at 340 K.\cite{Okuyama, Wilde} $\Delta d_{12}$ for
 $T_S$ higher than $T_D$, is +0.6$\sim$0.9 \%, which is in good
 agreement with that found for the $rapidly$ cleaned Pd(001). From the
 agreement of the atomic structures of the two differently prepared clean
 Pd(001), we conclude that clean Pd(001) must have almost the
 bulk-terminated structure, while hydrogen-adsorbed Pd(001) shows notable
  expansion of the $d_{12}$. \par

  \begin{table}
  \caption{ After adsorbing 1 ML of hydrogen on Pd(001) at 150 K,
 temperature of the sample is gradually raised to 430 K, during which
 LEED I/V spectra are taken and analyzed. {\it Hollow-H} signifies a model,
 1 ML of hydrogen adsorbed the four-fold hollow site of Pd(001), and
 $Clean$ does clean Pd(001).}
  \begin{ruledtabular}
  \begin{tabular}{cccccc}
  Exp. & \multicolumn{3}{c}{{\it Hollow-H}} & \multicolumn{2}{c}{{\it Clean}} \\
  \cline{2-4} \cline{5-6}
  temp. & $ d_{H}$ & $\Delta d_{12}/d_B$ & $R_P$-factor
     & $\Delta d_{12}/d_B$ & $R_P$-factor \\
  \tableline
  220 K & {\bf 0.20 \AA} &{\bf +4.5 \%} & {\bf 0.1692} & +5.3 \% & 0.2582 \\
  250 K & {\bf 0.20 \AA} &{\bf +4.4 \%} & {\bf 0.1748} & +5.2 \% & 0.2658 \\
  280 K & {\bf 0.21 \AA} &{\bf +4.4 \%} & {\bf 0.1826} & +5.2 \% & 0.2733 \\
  310 K & {\bf 0.21 \AA} &{\bf +4.2 \%} & {\bf 0.2088} & +5.1 \% & 0.2931 \\
  \tableline
  340 K & 0.44 \AA & +0.5 \% & 0.2007 &{\bf +0.9 \%} &{\bf 0.2004} \\
  400 K & 0.33 \AA & +0.6 \% & 0.1956 &{\bf +0.6 \%} &{\bf 0.1873} \\
  430 K & 0.61 \AA & +0.7 \% & 0.2043 &{\bf +0.8 \%} &{\bf 0.1899} \\
  \end{tabular}
  \end{ruledtabular}
  \end{table}


  \begin{figure}
  \includegraphics[width=0.45\textwidth]{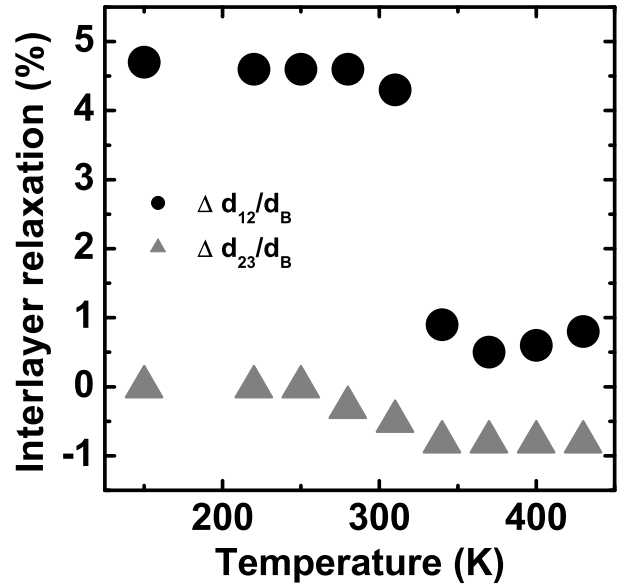}
  \caption{Dependence of $\Delta d_{12}$ and $\Delta d_{23}$ on sample
  temperature.}
  \end{figure}

  Although clean Pd(001) found in the present experiment show little surface
 relaxation, previous experiments report the expansion of $d_{12}$ by 2.5 \%
 to 3 \%.\cite{Behm0,Jona} The observation of the expansion of $d_{12}$ upon
 hydrogen adsorption suggests that the previous experiments were performed
 with hydrogen-contaminated Pd(001). For the direct examination of such
 possibility, we perform the sample preparation and the I/V acquisition in the
 $standard$ way without making extra-efforts such as flash cleaning and
 rapid cooling as made in the $rapid$ cleaning procedure. This experiment
 may correspond to a replication of previous experiments.\par

 In Table IV, given is a summary of I/V analyses for Pd(001) cleaned via
 the $standard$ procedure. The most notable result is that $d_{12}$
 increases by 1.8 \% (Also, see Fig. 5), which clearly contrasts to the almost
 bulk-terminated structure of clean Pd(001). $d_{12}$ in the present
 experiment, however, expands less than in previous reports, 2.5 \% to
 3 \%,\cite{Behm0,Jona} The reason why different amount of expansion of
 $d_{12}$ is observed experiment by experiment is supposed to be different
 degree of hydrogen contamination resulting from different base pressure and
 duration of experiment. To directly examine this conjecture, we dose
 varying amounts of hydrogen on Pd(001) at {\it room temperature}, and
 investigate the dependence of the atomic structure on hydrogen dosage by
 LEED I/V analysis. \par

  \begin{table}
  \caption{ LEED I/V analyses are made for Pd(001) cleaned via $standard$
 procedure, and for 6 ($6L-H$) and 12 L ($12L-H$) of hydrogen-dosed Pd(001).
 Respective I/V is fit, assuming clean Pd(001) to compare with previous
 results.}
 \begin{ruledtabular}
 \begin{tabular}{cccccc}
  & $\Delta d_{12}/d_{B}$ & $\Delta d_{23}/d_{B}$ & $R_P$-factor \\
  \tableline
  $standard$ & +1.8 \% & -0.7 \% & 0.1944 \\
  $6L - H$ & +2.4 \% & -0.7 \% & 0.2095 \\
  $12L- H$  & +3.0 \% & -1.0 \% & 0.2235 \\
  \end{tabular}
 \end{ruledtabular}
 \end{table}


  \begin{figure}
  \includegraphics[width=0.45\textwidth]{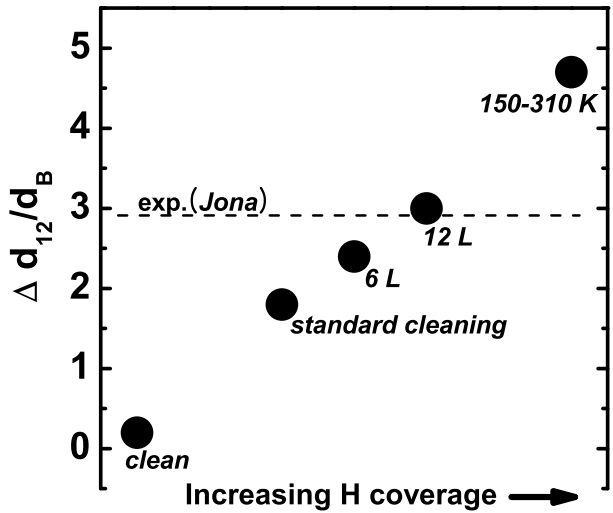}
  \caption{ $\Delta d_{12}/d_{B}$ is presented for increasing hydrogen
 coverage. The x-axis just indicates the direction of increasing hydrogen
 coverage, but is not in proportion. $clean$ signifies clean Pd(001) prepared
 by $rapid$ cleaning (Table I). {\it 150-310 K} refers to mean $\Delta
 d_{12}$ of 1 ML of hydrogen-covered surfaces from 150 K to 310 K.
 (Table III) {\it Standard cleaning} indicates Pd(001) cleaned via $standard$
 procedure, and $6 L$ and $12 L$ do Pd(001) dosed with the respective
 amount of hydrogen at room temperature. $Jona$ refers to Jona et. al.'s
 experiment.\cite{Jona}}
  \end{figure}

  In Table IV, given are the best-fit structures of Pd(001) 1) prepared via
 $standard$ cleaning and 2) dosed respectively by 6 and 12 L of hydrogen
 at room temperature. When 6 L of hydrogen is dosed, $d_{12}$ expands
 by 2.4 \%, while it does by 3.0 \% with increased dosage of 12 L. In regards
 to the fact that $d_{12}$ expands by 4.7 \% for hydrogen saturated surface
 at 150 K, we find monotonic increase of $\Delta d_{12}$ with increasing
 amount of hydrogen coverage.(Fig. 5) From the above account, the reason
 why the amount of expansion of $d_{12}$ varies experiment by experiment
 is evidently the different degree of hydrogen contamination due to different
 sample cleaning procedure and/or experimental environment. In short, the
 previous experimetal results of expanded $d_{12}$ of clean Pd(001) is
 attributed to the contamination by residual hydrogen. \par

  \section {Summary}

 We prepare clean Pd(001) in two different ways, $rapid$ cleaning and
 hydrogen desorption, and consistently find by LEED I//V analysis that the
 atomic structure of clean Pd(001) is similar to bulk-terminated Pd(001) .
 (Fig.1 and 4) On the other hand, hydrogen covered surface shows monotonic
 increase of $d_{12}$ with increasing hydrogen coverage. (Fig. 5) The
 present, systematic study on the effect of hydrogen on the atomic structure
 of Pd(001) firmly conclude that the previously reported expansion of
 interlayer spacing of Pd(001) originates from hydrogen contamination of the
 sample. \par

  \begin{acknowledgments}

  \end{acknowledgments}

 \end{document}